\documentclass{PoS}

\title{Quenched penguins, the $\Delta I=1/2$ rule, and 
$\varepsilon'/\varepsilon$}

\ShortTitle{Quenched penguins}

\author{M. Golterman\\
Department of Physics and Astronomy,
 San Francisco State University, 
 1600 Holloway Ave, San Francisco, CA 94132, USA\\
        E-mail: \email{maarten@stars.sfsu.edu  } }

\author{\speaker{E. Pallante}\\
     Institute for Theoretical Physics, University of Groningen,
Nijenborgh 4, 9747 AG  Groningen, The Netherlands   \\
        E-mail: \email{e.pallante@rug.nl}}

\abstract{
The transformation properties  of
strong penguin operators under the action of the flavor group change
when they are considered as operators in (partially) quenched QCD
instead of the unquenched theory.
An ambiguity arises, which is parameterized by new low-energy constants in the 
effective theory describing non-leptonic kaon decays in the (partially)
quenched setting.
Here we summarize results of the analysis for the complete set of three-flavor
strong penguin operators, consisting of products of two left-handed flavor 
currents, or a left- and a right-handed current. 
Our results imply that (partially) quenched lattice computations of the 
$\Delta I=1/2$ rule and  $\epsilon'/\epsilon$ are both affected by ambiguities 
intrinsic to the use of the quenched approximation at leading order in the
chiral expansion.  The only exception
is the partially quenched case with three light sea quarks, consistent
with general expectations.
We also address the issue of quenched ambiguities in the case of an active charm, 
correcting and extending that in Phys. Rev. D 74, 
014509 (2006).
    }

\FullConference{XXIVth International Symposium on Lattice Field Theory\\
                July 23-28, 2006\\
                Tucson, Arizona, USA}

\begin{document}
\newcommand{\be}{\begin{equation}}
\newcommand{\ee}{\end{equation}}
\newcommand{\ba}{\begin{eqnarray}}
\newcommand{\ea}{\end{eqnarray}}

\newcommand{\cL}{{\cal L}}
\newcommand{\cM}{{\cal M}}
\newcommand{\Bt}{{\tilde B}}
\newcommand{\cO}{{\cal O}}
\newcommand{\cOt}{{\tilde\cO}}
\newcommand{\bt}{{\tilde\beta}}
\newcommand{\tr}{{\mbox{tr}\,}}
\newcommand{\str}{{\mbox{str}\,}}
\newcommand{\Exp}{{\mbox{exp}\,}}
\newcommand{\Mdot}{{\dot M}}
\newcommand{\Mbar}{{M_{VS}}}
\newcommand{\tb}{{\tilde\beta}}
\newcommand{\vp}{{\vec p}}
\newcommand{\hX}{{\hat X}}
\newcommand{\diag}{{\rm diag}}
\newcommand{\sbar}{{\overline{s}}}
\newcommand{\dbar}{{\overline{d}}}
\newcommand{\ubar}{{\overline{u}}}
\newcommand{\qbar}{{\overline{q}}}
\newcommand{\psibar}{{\overline{\psi}}}
\newcommand{\tu}{{\tilde u}}
\newcommand{\tub}{{\overline{\tu}}}
\newcommand{\td}{{\tilde d}}
\newcommand{\tdb}{{\overline{\td}}}
\newcommand{\ts}{{\tilde s}}
\newcommand{\tsb}{{\overline{\ts}}}
\newcommand{\ie}{{\it i.e.}}
\newcommand{\eg}{{\it e.g.}}
\newcommand{\cf}{{\it cf.}}
\newcommand{\etc}{{\it etc.}}
\newcommand{\Nh}{{\hat N}}
\newcommand{\Real}{{\rm Re}}
\newcommand{\Imag}{{\rm Im}}
\newcommand{\eps}{{\varepsilon}}

\section{A group theory exercise}

In the theory where the charm quark has been integrated out penguin 
operators play
an important role.  In particular, referring to a commonly used basis, the
left-left (LL) penguin operator $\cO_2$ plays an important role in the
$\Delta I=1/2$ rule, while the left-right (LR) operator $\cO_6$  gives a 
major contribution to $\varepsilon'/\varepsilon$ \cite{dghb}.

When one makes the transition from unquenched QCD to partially-quenched (PQ)
QCD, the theory is changed from the physical theory with three light quarks
to a theory with $K$ light valence quarks and $N$ light sea quarks.
Fully quenched QCD is the special case with $N=0$.  This implies that the
flavor symmetry group changes from the usual $SU(3)_L\times SU(3)_R$ to
the graded group $SU(K+N|K)_L\times SU(K+N|K)_R$ \cite{bgpq}.
In general, this implies that the classification of weak operators with respect
to the flavor symmetry group also changes.  In particular, what happens for
strong penguins is that the penguin operator which transformed as a 
component of one irreducible representation (irrep) of $SU(3)_L\times SU(3)_R$
(the octet representation) now splits into several parts, each 
transforming in a {\em different} representation of the PQ symmetry group.  
One of those is the
``natural" generalization of the original penguin operator to the PQ theory,
whereas the other transforms in a more complicated way under
$SU(K+N|K)_L\times SU(K+N|K)_R$.  We will refer to these two parts as
the ``singlet" and ``adjoint" operators, respectively -- for reasons that will
become clear in the next section.
The problem reduces to a group theory exercise; the one of decomposing a 
given operator in terms of irreducible representations of the partially 
quenched group. The task will be a little more complicated in the LL case, 
but conceptually identical to the LR case and with similar results.  

\section{\large\bf Left-Right penguins in partially quenched QCD}

We consider LR penguin operators of the form~\cite{gp2,gp1}
\be
\cO_{penguin}=(\sbar d)_L(\ubar u+\dbar d+\sbar s)_R\, ,
\label{penguinLR}
\ee
where 
\ba
(\qbar_1 q_2)_{L,R}&=&\qbar_1\gamma_\mu P_{L,R}q_2\ ,\label{def}\\
P_{L,R}&=&\frac{1}{2}(1\mp\gamma_5) \ ,\nonumber
\ea
and color contractions are not specified, so that $\cO_{penguin}$ can 
represent both the color-mixed and un-mixed QCD penguins $Q_5$ and $Q_6$ (see \eg\   ref.~\cite{buras}).  
When we consider the LR penguin operator of Eq.~(\ref{penguinLR}) in the 
partially quenched theory, its representation content changes.  
A general realization of PQ QCD contains $K$ valence
quarks, each accompanied by one of $K$ ghost quarks with the same mass
in order to cancel the valence-quark determinant, and $N$ sea quarks -- the dynamical quarks -- which can all have masses
different from those of the valence quarks.   The relevant flavor symmetry 
group enlarges from the physical $SU(3)_L\times SU(3)_R$ to the graded group 
$SU(K+N|K)_L\times SU(K+N|K)_R$ \cite{bgpq}.\footnote{
For a detailed analysis of the actual symmetry group in the euclidean lattice theory, we refer
to Ref.~\cite{dgs}.  The upshot is that for our purposes, it is appropriate to
consider the PQ symmetry group to be $SU(K+N|K)_L\times SU(K+N|K)_R$
\cite{shsh}.}
It is clear that the $(\sbar d)_L$ factor in 
Eq.~(\ref{penguinLR})  is still a component of the
adjoint representation of $SU(K+N|K)_L$,
while the factor $(\ubar u+\dbar d+\sbar s)_R$ no longer
transforms as a singlet of $SU(K+N|K)_R$ . 
Instead, the operator can now be written as
\ba
\cO_{penguin}&=&
\frac{K}{N}\;
(\qbar\Lambda q)_L (\qbar q)_R + (\qbar\Lambda q)_L (\qbar A q)_R
 \nonumber \\
&\equiv&\frac{K}{N}\;\cO^{PQS} +\cO^{PQA}\ ,
\label{pqdecomp}
\ea
where we introduced the spurion fields $\Lambda$ and $A$ with values
\ba
\label{spurions}
\Lambda_i^{\ j}&=&\delta_{i3}\delta^{j2}\ ,\\
A&=&\diag\;\left(1-\frac{K}{N},\dots,-\frac{K}{N},\dots\right)\ .\nonumber
\ea
The first $K$ diagonal elements of $A$ are equal to $(1-K/{N})$ 
-- corresponding to the $K$ valence quarks -- and the last $N+K$ diagonal 
elements are equal to $(-K/N)$ -- corresponding to the $N$ sea quarks 
and the $K$ ghost quarks, both of which do not occur in $\cO_{penguin}$.  
The quark fields are graded vectors in flavor space, with fermionic components 
given by the valence and sea quarks, and bosonic components by ghost quarks.
The indices $i$ and $j$ are graded flavor indices, and run over valence, sea 
and ghost flavors.
For the down (strange) quark we have $i=2$ ($i=3$).  Notice that the decomposition of Eq.~(\ref{pqdecomp}) is singular in  the completely quenched 
theory, \ie, the theory with $N=0$.
For $N=K=3$ we regain the physical three-flavor 
theory. 

The spurions $\Lambda$ and $A$ both transform in the adjoint representation 
of $SU(K+N|K)$, as can be seen from the fact that both have a vanishing 
supertrace (str) \cite{bb}.
The operator $\cO^{PQS}$ thus transforms in the $(\mbox{adjoint}_L,1_R)$, 
while the operator $\cO^{PQA}$ transforms in the 
$(\mbox{adjoint}_L,\mbox{adjoint}_R)$. The appearance of the latter operator
is an artifact of the 
partially quenched setting.
The adjoint operator $\cO^{PQA}$ now contains only terms involving 
either sea or ghost quarks, and it is rather straightforward to see that 
their contributions to physical matrix elements
(\ie, those with only valence quarks on the external lines) vanishes because of
cancellation between sea-quark and ghost-quark loops. For this cancellation to
happen, valence masses and sea masses should be chosen equal.

In order to disentangle how this operator ambiguity affects kaon matrix 
elements, the effective low-energy realization of $\cO^{PQS}$ and $\cO^{PQA}$ 
is needed.
The bosonization of $\cO^{PQS}$ and $\cO^{PQA}$ leads straightforwardly to the 
following operators appearing at leading order in ChPT
\ba
\cO^{PQS}&\rightarrow& -\alpha^{(8,1)}_1\;\str(\Lambda L_\mu L_\mu)
+\alpha^{(8,1)}_2\;\str(\Lambda X_+)\ , \label{pqs} \\
\cO^{PQA}&\rightarrow&f^2\;\alpha^{(8,8)}\;
\str(\Lambda\Sigma A\Sigma^\dagger)
\ , \label{pqa}
\ea
where
\be
L_\mu=i\Sigma\partial_\mu\Sigma^\dagger\ ,\ \ \ \ \
X_\pm=2B_0(\Sigma M^\dagger\pm M\Sigma^\dagger)\ ,
\label{bb}
\ee
with $M$ the quark-mass matrix, $B_0$ the parameter $B_0$ of ref.~\cite{gl},
$\Sigma=\Exp(2i\Phi/f)$ the unitary field describing the partially quenched
Goldstone-meson multiplet, and $f$ the bare pion-decay constant
normalized such that $f_\pi=132$~MeV.  The $\alpha$'s are the corresponding
LECs.  
A striking result is that $\cO^{PQA}$,
unlike $\cO^{PQS}$, is of order $p^0$, due to the fact that
the right-handed current in $\cO^{PQA}$
is not a partially quenched singlet (\cf\ electro-magnetic
penguins\footnote{In fact, $\cO^{PQA}$ is a component of the same
irrep as the electro-magnetic penguin, except for $N=0$ \cite{gp2}.}).  
However, the new operator
$\cO^{PQA}$ does not contribute at tree level to matrix elements
with only valence quarks on external lines, since the matrix $A$
is effectively proportional  to the unit matrix 
in the valence sector.  
This is no longer true at next-to-leading-order, \ie, at order 
$p^2$, where one-loop contributions from $\cO^{PQA}$ to valence-quark 
matrix elements are non-zero and of the same chiral order as the leading 
(tree-level)
contribution from $\cO^{PQS}$ \cite{gp2}.  In the quenched case, the low-energy
constant $\alpha^{(8,8)}$ (which for $N=0$ is denoted as $\alpha^{NS}$) has
found to be numerically large \cite{golper,Laiho}.

\section{\large\bf Left-Left penguins in partially quenched QCD}
%
%
%
The other type of strong penguin operators is made by the product of two 
left-handed currents:
\ba
\label{penguins}
\cO_1&=&(\sbar d)_L (\ubar u)_L - (\sbar u)_L (\ubar d)_L\\
&=&(\sbar_\alpha d_\alpha)_L (\ubar_\beta u_\beta+\dbar_\beta d_\beta+\sbar_\beta s_\beta)_L -
(\sbar_\alpha d_\beta)_L (\ubar_\beta u_\alpha+\dbar_\alpha d_\beta+\sbar_\beta s_\alpha)_L\ ,
\nonumber\\
 \cO_2&=&(\sbar d)_L (\ubar u)_L + (\sbar u)_L (\ubar d)_L
 +2(\sbar d)_L (\dbar d+\sbar s)_L\nonumber\\
&=&(\sbar_\alpha d_\alpha)_L (\ubar_\beta u_\beta+\dbar_\beta d_\beta+\sbar_\beta s_\beta)_L +
(\sbar_\alpha d_\beta)_L (\ubar_\beta u_\alpha+\dbar_\alpha d_\beta+\sbar_\beta s_\alpha)_L\,  .
\nonumber
\ea
We have made the color indices $\alpha,\beta$ explicit, where needed.  
Both operators $\cO_{1,2}$ are linear combinations
of color un-mixed and color mixed terms, and transform in the octet 
representation of $SU(3)_L$, while, trivially, in the singlet representation of $SU(3)_R$.
Together with the LR operator in Eq.~(\ref{penguinLR}), they are part of a 
basis of irreducible representations of the chiral 
group that are CPS~\cite{cbetal} invariant and with definite isospin $I=1/2$ and 
$I=3/2$~\cite{dghb}. This basis is especially convenient for working out 
group theoretical properties. The relation with the frequently used basis of 
the $Q_i$ operators \cite{buras} can be found in Appendix A of Ref.~\cite{GPLL}. 
Within $SU(K+N|K)_L\times SU(K+N|K)_R$ the LL operators can now be written as
\ba
\label{split}
\cO_1&=&\frac{K}{N}\;\cO_-^{PQS}+\cO_-^{PQA}\ ,\\
\cO_2&=&\frac{K}{N}\;\cO_+^{PQS}+\cO_+^{PQA}\ ,\nonumber\\
&&\nonumber\\
\cO_\pm^{PQS}&=&(\qbar_\alpha\Lambda q_\alpha)_L (\qbar _\beta q_\beta)_L
\pm (\qbar_\alpha\Lambda q_\beta)_L (\qbar _\beta q_\alpha)_L\ ,\nonumber\\
\cO_\pm^{PQA}&=&(\qbar_\alpha\Lambda q_\alpha)_L (\qbar _\beta A q_\beta)_L
\pm (\qbar_\alpha\Lambda q_\beta)_L (\qbar _\beta A q_\alpha)_L\, .\nonumber
\ea
This time the operators $\cO_\pm^{PQA}$
transform as the product representation of two adjoint irreps, and they are 
thus reducible. 

The corresponding decomposition of $\cO_\pm^{PQA}$ is accomplished by (anti-)symmetrization
in covariant and contravariant indices, and by ``removing" supertraces on 
pairs of covariant and contravariant indices, much as is done in the case of $SU(N)$ \cite{bb}.
Here we take the quark fields $q_i$ as covariant, and the anti-quark fields
$\qbar^i$ as contravariant.  It turns out that the operators $\cO^{PQA}_-$ and
$\cO^{PQS}_-$ ($\cO^{PQA}_+$ and $\cO^{PQS}_+$)
are already symmetric (anti-symmetric) in both their two covariant and their two contravariant flavor indices -- see Ref.~\cite{GPLL} for details.  
A supertraceless linear combination exists, given by 
$\cO^{PQA}_\pm +{2}/{(\mp N-2)}\;\cO^{PQT}_\pm$, with $\cO^{PQT}_\pm=
(\qbar_\alpha\Lambda A q_\alpha)_L (\qbar _\beta  q_\beta)_L
\pm (\qbar_\alpha\Lambda  A q_\beta)_L (\qbar _\beta  q_\alpha)_L$. The singularity in the decomposition into irreps of $\cO^{PQA}_-$ for $N=2$, tells that 
the representation in which $\cO^{PQA}_-$ transforms is not fully reducible for this value of the number of sea quarks $N$. The bosonization of $\cO^{PQT}_\pm$
leads to the term $\cL_3^A$ in the lagrangian of Eq.~(\ref{bosonization}) below.

\section{The effective lagrangian}

The construction of the low-energy bosonized effective lagrangian for the
 operators $\cO_{1,2}$ in the PQ theory follows the same lines as the LR case. 
The less straightforward aspects are related to the decomposition of the 
operators in terms of irreps of the PQ group. To each independent 
representation of the group is associated a low-energy constant in the 
effective lagrangian. To lowest order in the chiral
expansion, with the $\eta'$ integrated out (and making use of CPS invariance) 
\be
\label{bosonization}
\cO_\pm^{PQA}\to\cL^A_\pm= \alpha^{A\pm}_{1a}(\cL_1^A\pm\cL_2^A)
+\alpha^{A\pm}_{1b}\cL_3^A+\alpha^{A\pm}_2\cL_4^A\  ,
\ee
with 
\ba
\label{effops}
\cL_1^A&=&\str(\Lambda L_\mu)\;\str(AL_\mu)\ ,\\
\cL_2^A&=&\str(\Lambda L_\mu AL_\mu)\ ,\nonumber\\
\cL_3^A&=&\str(\Lambda AL_\mu L_\mu)\ ,\nonumber\\
\cL_4^A&=&\str(\Lambda AX_+)\ .\nonumber
\ea
The bosonization rules that lead to Eq.~(\ref{bosonization}) rely on the decomposition of $\cO_\pm^{PQA}$ into irreps. Details can be found in Ref.~\cite{GPLL}.
We have explicitly indicated the dependence of the LECs on the
operator through the superscripts $\pm$, because they refer to different
representations of the PQ flavor group.  We conclude that the transition from
the unquenched theory to the PQ theory leads to the introduction of three
new LECs for each of the two operators $\cO_1$ and $\cO_2$.
These operators do contribute already at leading chiral order to $K^0\to vacuum$, $K^+\to \pi^+$ and $K\to 2\pi$ matrix elements. 
We also find that the one-loop corrections
for the adjoint operators, calculated in Ref.~\cite{GPLL}, differ from those of the singlet operators,
calculated in Ref.~\cite{gp1}. In other words, the singlet and adjoint LECs do
not occur in some fixed, given linear combinations in physical matrix 
elements beyond tree level.  Our conclusions for the fully quenched case ($N=0$),
for which the relevant symmetry group is $SU(K|K)$ are very 
similar, although the group-theoretical details are different from those of the
PQ ($N\ne 0$) case \cite{gp2,GPLL}.

\section{The $\Delta I=1/2$ rule and $\varepsilon'/\varepsilon$}

Our results show that quenching ambiguities do affect, already at the leading 
chiral order, those $\Delta S=1$ weak 
matrix elements that receive contributions from LR and LL penguin operators.
We emphasize that the new adjoint operators $\cO^{PQA}$ and $\cO^{PQA}_\pm$ 
occurring in the PQ theory are genuinely new operators, and one thus expects that one-loop corrections in ChPT for matrix elements of these operators 
differ from those of the singlet operators. We find that this is indeed the 
case \cite{GPLL}.
In the case of LR penguins, the enhancement of the adjoint
operators leads to the appearance of chiral logarithms already at leading order
in ChPT \cite{gp2}.

The quenching ambiguity affecting the LR penguin operator $Q_6$ has 
dramatic consequences for the quenched lattice determination of $\varepsilon '/\varepsilon$~\cite{Japan}, 
and may provide an explanation for earlier quenched lattice results~\cite{cppacs,rbc}.
The quenching ambiguity in LL penguin operators also directly affects the 
$\Delta I=1/2$ rule for which the dominant 
contributions come from the current-current operators $Q_1$ and $Q_2$, where 
$Q_1={1}/{2}\cO_1+{1}/{10}\cO_2+{1}/{15}\cO_3+{1}/{3}\cO_4$ and 
$Q_2=-{1}/{2}\cO_1+{1}/{10}\cO_2+{1}/{15}\cO_3+{1}/{3}\cO_4$.\footnote{We 
disagree with statements to the contrary about the $\Delta I=1/2$ amplitude
in Ref.~\cite{Laiho}.}
In addition, quenching ambiguities in LL penguin operators can in principle also
affect a lattice determination of $\varepsilon '/\varepsilon$ through the operator $Q_4$  
because $Q_4=-{1}/{2}\cO_1+{1}/{2}\cO_2$.  This 
can be relevant in the presence of a large cancellation of the dominant contributions from $Q_6$ and the electroweak penguin operator 
$Q_8$~\cite{Japan}.

\section{The theory with active charm}

A natural setting for a lattice investigation of non-leptonic kaon weak 
matrix elements is the three-flavor effective theory, where the light quarks 
$u,\, d,\, s$ are kept dynamical. However, the lattice implementation of a 
4-flavor theory, where the charm is active at its physical mass, has various 
advantages:
less operator mixing, the GIM mechanism is at work, and the short- to long-distance matching 
scale $\mu >m_c$ is comfortably high for a perturbative RG evolution of the 
Wilson coefficients.
It is thus relevant to understand how (partial) quenching modifies the 4-flavor theory, and more specifically, if and how the quenching
 ambiguities of the 3-flavor penguin operators survive in the 4-flavor theory.
What matters for the appearance of the ambiguity is the existence of
 penguin operators in the first place. This is a consequence of the fact that 
what matters are the transformation properties of the given operators 
under the flavor chiral group. Again, it is a group theory exercise: 
it  does not depend on the relative energy scales involved, nor 
on the largeness of chiral symmetry breaking induced by mass terms.
Once the charm is active, the chiral flavor group is
$SU(4)_L\times SU(4)_R$, and one needs to consider the classification of
the weak effective hamiltonian under (the PQ generalization of) this group.

At this point the way penguins enter in physical matrix elements is different for the CP conserving 
$\Delta I=1/2$ rule and for the CP violating parameter 
$\varepsilon'/\varepsilon$.  
The GIM mechanism ensures that the Wilson coefficients $z_i(\mu )$ 
-- \ie\ those contributing to CP conserving amplitudes -- of penguin 
operators\footnote{We limit the discussion to operators at order $G_F$.}
are zero in the five- and four-flavor theory, \ie, until the threshold 
$\mu\leq m_c$
is crossed. As a result, penguins in $SU(4)$ do not contribute to the 
$\Delta I=1/2$ ratio, and this leads to the conclusion that no quenched 
ambiguity arises for this case.\footnote{The quenching ambiguity affects 
operators containing bilinears which transform as singlets under the flavor group. The operators considered in Ref.~\cite{GPLL} in the case of active charm do not contain bilinears in singlets of $SU(4)$. } This conclusion corrects the observation made by the authors 
in Ref.~\cite{GPLL}. 

The situation is different in the case of $\varepsilon'/\varepsilon$, where the
penguin operators $Q_i,\ i=3,4,5,6$ (see \eg\ Ref.~\cite{buras} for a derivation of the effective $\Delta S=1$ hamiltonian in the four-flavor theory) 
do contribute to  $\varepsilon'/\varepsilon$.
The operators $Q_i,\ i=3,4,5,6$ are now the $SU(4)$ extension of their $SU(3)$ counterparts, 
\ie, now $\sum \bar{q}q=\bar{u}u+\bar{d}d+\bar{s}s+\bar{c}c$. The way 
(partial) quenching effects (the singlet bilinears in) penguin operators in 
$SU(4)$ is analogous to the $SU(3)$ case. It must be concluded that a quenching 
ambiguity does effect LR and LL penguin contributions to 
$\varepsilon'/\varepsilon$ also in the four-flavor theory with an active charm quark.

\section*{Acknowledgments}

 We thank Pilar Hern\'andez for a useful discussion on the role of 
active charm in the $\Delta I=1/2$ rule, which motivated the discussion
contained in this paper.   This corrects and extends the observation made in 
Ref.~\cite{GPLL}.
MG was supported in part by the Generalitat de Catalunya under the program
PIV1-2005 and by the US Department of Energy. EP is supported by the 
University of Groningen.

\end{document}